\documentclass[letter,twocolumn]{jpsj3} 
\title{Numerical-Diagonalization Study of Spin Gap Issue \\ 
of the Kagome Lattice Heisenberg Antiferromagnet}
\catcode`\@=11
\def\simle{\mathrel{\mathpalette\@versim<}}   
\def\simge{\mathrel{\mathpalette\@versim>}}   
\def\@versim#1#2{\lower2.5pt\vbox{\baselineskip0pt \lineskip-.5pt
   \ialign{$\m@th#1\hfil##\hfil$\crcr#2\crcr\sim\crcr}}}
\catcode`\@=12

\author{Hiroki \textsc{Nakano}
\thanks{E-mail address: hnakano@sci.u-hyogo.ac.jp}
and 
Toru \textsc{Sakai}$^{1}$
}

\inst{Graduate School of Material Science, University of Hyogo,
Kouto 3-2-1, Kamigori, Ako-gun, Hyogo 678-1297, Japan \\
$^{1}$
Japan Atomic Energy Agency, SPring-8, 
Kouto 1-1-1, Sayo, Hyogo 679-5148, Japan
}

\recdate{\today}

\abst{We study the system size dependence of 
the singlet-triplet excitation gap 
in the $S=1/2$ kagome-lattice Heisenberg antiferromagnet 
by numerical diagonalization. 
We successfully obtain a new result of a cluster 
of 42 sites. 
The two sequences of gaps of systems 
with even-number sites and that with odd-number sites 
are separately analyzed. 
Careful examination clarifies that 
there is no contradiction 
when we consider the system to be gapless. 
}

\kword{antiferromagnetic Heisenberg spin model,
kagome lattice, frustration, 
numerical-diagonalization method, Lanczos method}

\begin{document}
\maketitle

The kagome-lattice antiferromagnet attracts 
the attention of many researchers 
because it is one typical case in which 
magnetic frustration occurs in systems. 
Recently, the discovery of candidate materials, i.e. , 
herbersmithite, volborthite, and vesignieite, 
has promoted further investigation 
into clarifying their properties\cite{Mendels}. 
However, the kagome-lattice antiferromagnet 
is still an unresolved system, 
some properties of which remain unclear.  

One of them is the spin gap issue 
of the $S=1/2$ kagome-lattice Heisenberg antiferromagnet,  
namely, whether or not a nonzero energy gap 
exists between the ground-state singlet level 
and the lowest triplet one. 
The reason why this issue is difficult to resolve is that 
some of the numerical methods are not adequate at present. 
The quantum Monte Carlo (QMC) simulation has a so-called 
negative sign problem due to the frustration.  
The density matrix renormalization group (DMRG) method 
is not good because of the two-dimensionality of the system 
although much effort 
using the DMRG method\cite{Yan_Huse_White_DMRG} has recently  
been carried out, which will be mentioned later. 
Under such circumstances, 
numerical diagonalization calculation can provide us with 
precise data; however, available system sizes are 
limited to being very small.  
In studies of the kagome-lattice antiferromagnet, 
only sizes up to 36 sites were 
treated\cite{Lecheminant,Waldtmann,Hida_kagome,Cepas,
Cabra,Honecker0,Honecker1,Spin_gap_Sindzingre,HN_Sakai2010} 
before Sakai and Nakano treated a system with 39 sites 
for the first time\cite{Sakai_HN_PRBR}. 
Although Sindzingre and Lhuillier\cite{Spin_gap_Sindzingre} 
attempted to resolve the spin gap issue 
in the argument on the basis of the data 
up to 36 sites, the authors finally concluded 
``it is impossible to distinguish between a gapless system 
and a system with a very small gap.'' 
Therefore, numerical results of larger system sizes 
are required in order to tackle this issue; 
thus, carrying out large-scale numerical-diagonalization 
calculations is an urgent task. 

The purpose of this study is to investigate 
the spin gap issue from the argument 
based on new numerical-diagonalization data 
of larger systems that have not yet been reported. 
We successfully obtain some new results in the case 
of up to 42 sites. 

Here, we examine the excitation gap 
of the $S=1/2$ kagome-lattice Heisenberg antiferromagnet. 
The Hamiltonian is given by 
\begin{equation}
{\cal H} = \sum_{\langle i,j\rangle} J
\mbox{\boldmath $S$}_{i}\cdot\mbox{\boldmath $S$}_{j},
\label{H_kagome}
\end{equation}
where $\mbox{\boldmath $S$}_{i}$ 
denotes the $S=1/2$ spin operator. 
The sum runs over all the nearest-neighbor pairs. 
Energies are measured in units of $J$; hereafter, we set $J=1$.  
The number of spin sites is denoted by $N_{\rm s}$, 
where $N_{\rm s}/3$ is an integer. 
We impose the periodic boundary condition 
for clusters with site $N_{\rm s}$. 
We calculate the lowest energy of ${\cal H}$ 
in the subspace divided by $\sum _j S_j^z=M$. 
The energy is denoted by $E(N_{\rm s},M)$. 
\begin{figure}[htb]
\begin{center}
\includegraphics[width=7cm]{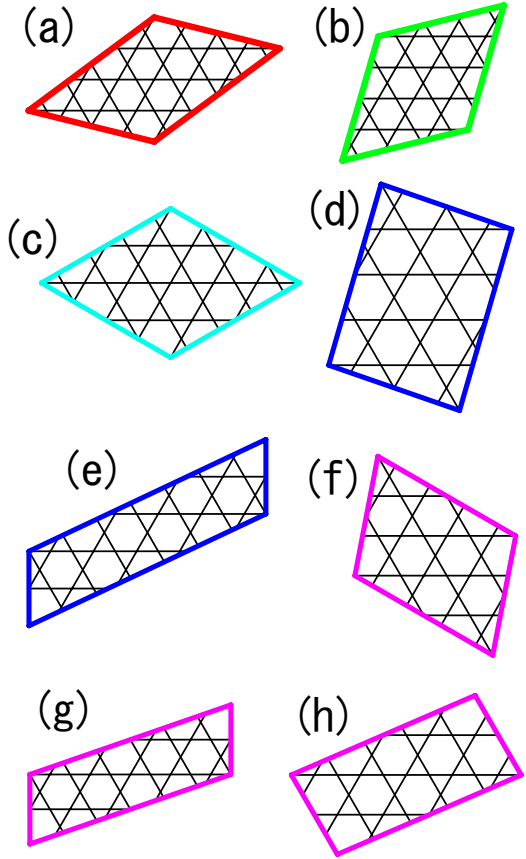}
\end{center}
\caption{Shapes of finite-size clusters for 
(a) $N_{\rm s}=42$, 
(b) $N_{\rm s}=39$, 
(c) $N_{\rm s}=36$, 
(d) $N_{\rm s}=33$, 
(e) $N_{\rm s}=33$,   
(f) $N_{\rm s}=30$, 
(g) $N_{\rm s}=30$, and 
(h) $N_{\rm s}=30$.  
Note that clusters (b) and (c) are rhombic 
and the others are not. 
}
\label{fig1}
\end{figure}

Using the numerical exact diagonalization 
based on the Lanczos algorithm, 
we have calculated 
the values of $E(N_{\rm s},M)$ of clusters up to 
$N_{\rm s}=42$, 
to obtain the finite-size energy differences. 
The convergence of our Lanczos calculations 
has been checked so that the relative changes 
of energies of the lowest and first-excited states 
become less than $10^{-10}$. 
The shapes of the finite-size clusters 
of system sizes from $N_{\rm s}=30$ to $N_{\rm s}=42$ 
are shown in Fig.~\ref{fig1}. 
For $N_{\rm s}=36$ and 39, 
a rhombic cluster having an interior angle $\pi/3$ 
is available, in which two-dimensionality 
may be captured well. 
For other system sizes of $N_{\rm s}=30$, 33, and 42, 
on the other hand, no rhombic cluster is available. 
For $N_{\rm s}=30$ and 33 in such cases, 
we examine more shapes than one specific case; 
we here present different results due to the difference 
between cluster shapes as many as we have obtained. 
Since the difference between cluster shapes 
is regarded as a kind 
of boundary conditions, the effect on our results for the gaps 
appears as a finite-size effect due to such various boundaries. 
In our analysis, it is desirable 
to extract information that is not 
largely affected from a specific cluster shape.  
To realize such an analysis, 
we treat various results from different nonrhombic shapes 
equivalently instead of rejecting the results 
in nonrhombic cases. 
The case of $N_{\rm s}=42$ shows the largest size in this study; 
thus, we treat only a single case of shape 
owing to the limitation 
of available computational time that we have prepared. 
For $N_{\rm s}=42$, the largest dimension is 538~257~874~440 
for the subspace of $M=0$. 
To treat such huge matrices in computers, we have carried out 
parallel calculations\cite{comm_performance} 
using the MPI-parallelized code\cite{comments2para}. 
Note that this dimension is the largest 
among numerical-diagonalization studies 
of quantum lattice systems, 
as far as we know\cite{comment_dim}. 
The ground-state energy per site 
for a given $N_{\rm s}$ is given by
\begin{equation}
e_{g}=
\left\{
\begin{array}{ll}
E(N_{\rm s},M=0          )/N_{\rm s} & (\mbox{even}\ N_{\rm s} ) 
\\
E(N_{\rm s},M=\frac{1}{2})/N_{\rm s} & (\mbox{odd}\ N_{\rm s} ) .
\end{array}
\right.
\end{equation}
We evaluate the finite-size energy difference as 
\begin{equation}
\Delta_{N_{\rm s}}=
\left\{
\begin{array}{ll}
E(N_{\rm s},M=1)-E(N_{\rm s},M=0) & (\mbox{even}\ N_{\rm s} ) 
\\
E(N_{\rm s},M=\frac{3}{2})-E(N_{\rm s},M=\frac{1}{2}) 
& (\mbox{odd}\ N_{\rm s} ) .
\end{array}
\right.
\end{equation}
Note here that 
$\Delta_{N_{\rm s}}$ for even $N_{\rm s}$ and 
$\Delta_{N_{\rm s}}$ for odd $N_{\rm s}$ 
are supposed to be the same in the thermodynamic limit, 
whereas the two kinds of $\Delta_{N_{\rm s}}$ are 
different from each other when $N_{\rm s}$ is finite. 
One should consider the difference 
for finite $N_{\rm s}$\cite{comment_Mendels}. 

\begin{figure}[htb]
\begin{center}
\includegraphics[width=9cm]{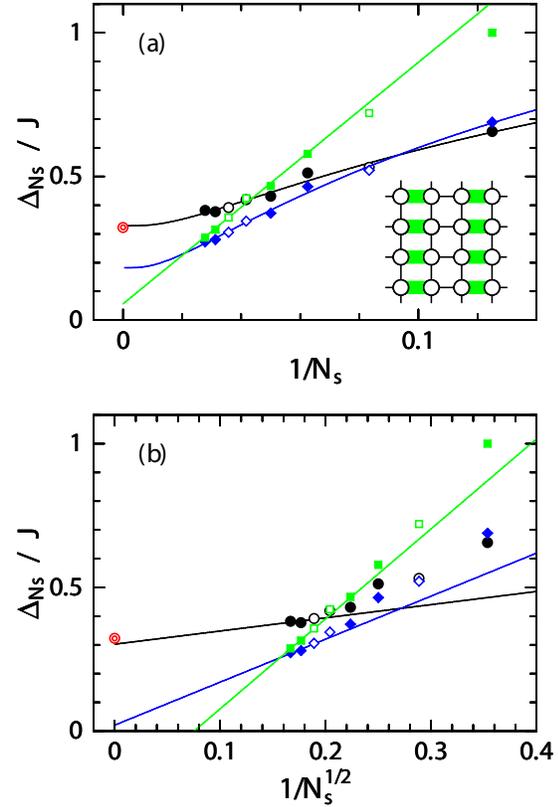}
\end{center}
\caption{Analysis of the spin gap 
of the Heisenberg antiferromagnet 
on the dimerized square lattice. 
The $1/N_{\rm s}$ 
and $1/N_{\rm s}^{1/2}$ dependences 
of $\Delta_{N_{\rm s}}/J$ 
are plotted in panels (a) and (b), respectively. 
The inset in panel (a) illustrates the lattice structure, 
in which thick green bonds and thin black bonds 
denote the interactions of $J$ and $J^{\prime}$, respectively. 
Green squares, blue diamonds, and black circles 
represent the cases of $\alpha$=1, 0.52337, and 0.4, 
respectively. 
Closed symbols mean that the shapes of the clusters 
are regular squares; open symbols mean that they are not. 
A red double circle in each panel denotes an estimate 
from the QMC simulation\cite{matsumoto_dimer_sq}. 
The explanations of fitting lines and curves are 
given in the text. 
}
\label{fig2}
\end{figure}

Before investigating the kagome-lattice antiferromagnet, 
we consider the case of the Heisenberg antiferromagnet 
on the dimerized square lattice 
illustrated in the inset of Fig.~\ref{fig2}(a) 
because reliable numerical estimates 
from the QMC simulations have been 
reported\cite{matsumoto_dimer_sq}.  
The Hamiltonian is given by 
\begin{equation}
{\cal H} = \sum_{\langle i,j\rangle} J
\mbox{\boldmath $S$}_{i}\cdot\mbox{\boldmath $S$}_{j}
+
\sum_{\langle i,j\rangle} J^{\prime}
\mbox{\boldmath $S$}_{i}\cdot\mbox{\boldmath $S$}_{j}. 
\label{H_dimer_sq}
\end{equation}
Here, the ratio of $J^{\prime}/J$ denoted by $\alpha$ 
is a controllable parameter. 
The case of $\alpha=1$ is that of a simple square lattice 
in which the ground state is N$\acute{\rm e}$el-ordered. 
On the other hand, the case of $\alpha=0$ is reduced 
to an assembly of isolated dimers; 
the gapped ground state is realized near $\alpha=0$.  
Matsumoto {\it et al}. clarified 
the boundary between these phases to be $\alpha=0.52337(3)$. 
The examination of this model makes it possible 
to confirm an appropriate means of analyzing 
$\Delta_{N_{\rm s}}$ of diagonalization data 
in two-dimensional systems; 
results are shown in Fig.~\ref{fig2}. 
In Fig.~\ref{fig2}(a), analysis based on the case 
when a nonzero gap exists is demonstrated; 
finite-size data are fitted by 
\begin{equation}
\label{fit4gapped}
\Delta_{N_{\rm s}}= \Delta_{\infty}
+C \exp(-D {N_{\rm s}^{1/2}} ), 
\end{equation} 
which was also used in ref.~\ref{matsumoto_dimer_sq}. 
One finds that the gapped case of $\alpha=0.4$ 
gives us the result of $\Delta_{\infty}$ 
that agrees well with the precise estimate 
from the QMC simulation\cite{matsumoto_dimer_sq}. 
However, the case of $\alpha=0.52337$ in which 
the system is gapless (and critical) incorrectly 
gives a nonzero $\Delta_{\infty}$.  
In the gapless case of $\alpha=1$, the fitting is a failure; 
even the linear line obtained by fitting data of systems 
larger than $N_{\rm s}=16$ intersects at the intercept of 
a positive value with the line of $N_{\rm s}\rightarrow\infty$. 

In Fig.~\ref{fig2}(b), on the other hand, 
analysis based on the gapless case is carried out; 
the fitting is based on the line 
\begin{equation}
\label{fit4gapless}
\Delta_{N_{\rm s}}= F +\frac{G}{N_{\rm s}^{1/2}}, 
\end{equation} 
which is related to the assumption that 
the system has a dispersion relation proportional to 
the wave number having the unit of the inverse length. 
The linear fitting of data $N_{\rm s}$=36, 32, and 28 
in the case of $\alpha=0.52337$ ($\alpha=1$) 
gives a negative (almost vanishing) $F$, 
which suggests that the system is gapless\cite{comm_negative}. 
In the gapped case of $\alpha=0.4$, 
a positive $F$ is very close to the precise estimate 
from the QMC simulation\cite{matsumoto_dimer_sq}. 
Therefore, these examinations of Figs.~\ref{fig2}(a) and (b) 
indicate that, in two-dimensional systems,  
the analysis based on Fig.~\ref{fig2}(a) is good at 
capturing gapped cases and 
the analysis based on Fig.~\ref{fig2}(b) 
is better for recognizing gaplessness. 


Now, we present our new results 
of the kagome-lattice antiferromagnet of $N_{\rm s}=42$ 
accompanied by the results of smaller clusters, 
some of which have already been reported; 
these results are shown in Table~\ref{t1}. 
\begin{table}[h]
\caption{List of numerical data of the ground state energy 
and the singlet-triplet energy difference. 
The last column shows the first report 
of $\Delta_{N_{\rm s}}/J$ for each case. }
\label{t1}
\begin{center}
\begin{tabular}{lllll}
\hline
\multicolumn{1}{c}{$N_{\rm s}$} & 
\multicolumn{1}{c}{shape} & 
\multicolumn{1}{c}{$-e_{\rm g}/J$} & 
\multicolumn{1}{c}{$\Delta_{N_{\rm s}}/J$} & 
\multicolumn{1}{c}{first report} 
\\
\hline
42 & Fig.~\ref{fig1}(a) & 0.4381425934 & 0.14909214 & new \\
39 & Fig.~\ref{fig1}(b) & 0.4364152017 & 0.22243392 & ref.~\ref{Sakai_HN_PRBR}\\
36 & Fig.~\ref{fig1}(c) & 0.4383765311 & 0.16420678 & ref.~\ref{Waldtmann}\\
33 & Fig.~\ref{fig1}(d) & 0.4366725772 & 0.22945504 & ref.~\ref{HN_Sakai2010}\\
33 & Fig.~\ref{fig1}(e) & 0.4430755530 & 0.25405132 & new \\

30 & Fig.~\ref{fig1}(f) & 0.4384772983 & 0.15285554 & ref.~\ref{Lecheminant} \\
30 & Fig.~\ref{fig1}(g) & 0.4458946475 & 0.16823452 & new \\
30 & Fig.~\ref{fig1}(h) & 0.4452556126 & 0.17565828 & ref.~\ref{Mendels} \\
\hline
\end{tabular}
\end{center}
\end{table}
Note that the present study is the first study, 
to the best of our knowledge, that presents 
the result of a singlet-triplet energy difference 
in the $S=1/2$ system with $N_{\rm s}=42$, 
whereas the ground-state energy of the system with $N_{\rm s}=42$ 
in several cases of the star lattice\cite{Honecker_star}, 
one-dimensional uniform chain\cite{spinpack_chain}, 
and kagome lattice\cite{comment_Lauchli} 
has been reported.  
Note that our results of the ground-state energy per site 
are very close to the recent estimate $e_{\rm g}/J=-0.4379(3)$ 
from the DMRG calculations\cite{Yan_Huse_White_DMRG}. 
In this study, we focus our attention 
on the singlet-triplet energy differences 
and their system size dependence hereafter. 

First, let us observe the results of $\Delta_{N_{\rm s}}/J$ 
for $N_{\rm s}=30$. 
The cases shown in Figs.~\ref{fig1}(f)-\ref{fig1}(h) 
depart from the ideal two-dimensional case
in different degrees. 
When one examines the degree of strip, 
the departure becomes stronger 
in the order of clusters (f), (h), and (g). 
However, $\Delta_{N_{\rm s}}/J$ 
becomes larger in the order of clusters (f), (g), and (h). 
The two orders disagree with each other. 
The disagreement indicates that, 
within the viewpoint of the results of $\Delta_{N_{\rm s}}/J$, 
determining whether the two-dimensionality is better or worse 
is not very trivial. 
Thus, this suggests that there is considerable significance 
in examining various cluster shapes in nonrhombic cases, 
irrespective of the degree of strip. 

\begin{figure}[htb]
\begin{center}
\includegraphics[width=11cm]{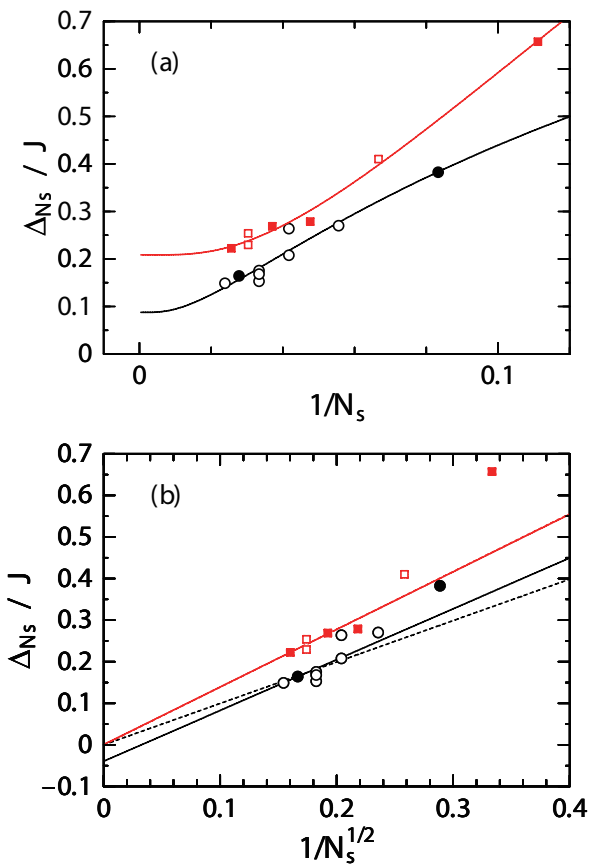}
\end{center}
\caption{Analysis of the spin gap 
of the Heisenberg antiferromagnet 
on the kagome lattice. 
The $1/N_{\rm s}$ 
and $1/N_{\rm s}^{1/2}$ dependences of $\Delta_{N_{\rm s}}/J$ 
are plotted in panels (a) and (b), respectively. 
Red squares (black circles) represent 
the case of odd (even) $N_{\rm s}$. 
Closed symbols mean that the clusters 
are rhombus with an interior angle $\pi/3$; 
open symbols mean that they are not. 
The explanations of the fitting lines and curves are 
given in the text. 
}
\label{fig3}
\end{figure}

Let us plot the $1/N_{\rm s}$ dependence of singlet-triplet gaps 
of the kagome-lattice antiferromagnet in Fig.~\ref{fig3}(a), 
where the data listed in Table~\ref{t1} and some other results 
for clusters with a smaller $N_{\rm s}$ are presented. 
One finds that the two sequences of even-$N_{\rm s}$ and 
odd-$N_{\rm s}$ data are different from each other. 
Thus, we analyze these two sequences separately 
on the basis of the assumption of the gapped case, 
as explained in Fig.~\ref{fig2}(a). 
We obtain $\Delta_{\infty}$=0.088(69) from the data 
of even $N_{\rm s}$ 
and $\Delta_{\infty}$=0.209(18) from the data 
of odd $N_{\rm s}$. 
The two extrapolated results disagree with each other 
even though the fitting errors are taken into account. 
The disagreement suggests that 
the analysis under the assumption of the gapped case is invalid. 
There are two possibilities for this invalidity. 
One is that the system is not gapped. 
The above nonzero estimates of the spin gap 
are artificial consequences from the invalid analysis. 
(See the case of $\alpha=0.52337$ in Fig.~\ref{fig2}(a).) 
The other possibility is that 
although the system is intrinsically gapped,  
the system size dependence is completely different 
from eq.~(\ref{fit4gapped}). 
The estimate $\Delta_{\infty}$=0.088(69) from even $N_{\rm s}$ 
coincides within the error 
with an estimate 0.05 of the singlet-singlet gap 
obtained from the recent DMRG calculations 
\cite{Yan_Huse_White_DMRG} 
as a lower bound for the singlet-triplet (spin) gap.  
However, $\Delta_{\infty}$=0.209(18) from odd $N_{\rm s}$ 
is not in agreement with 0.05 from the DMRG method. 
If the kagome-lattice antiferromagnet has 
a nonzero spin gap close to 0.05, 
this leads to the data sequence of finite-size gaps 
of odd $N_{\rm s}$, 
revealing an unknown and/or peculiar system size dependence. 
Even though one does not use eq.~(\ref{fit4gapped}), 
one finds that the data sequence of odd $N_{\rm s}$ 
does not show the behavior of approaching a value close to 0.05. 
Thus, we think that this possibility is quite small. 

Next, we carry out an analysis of the data 
in Fig.~\ref{fig3}(b), 
under the assumption in the gapless case in the manner 
explained in Fig.~\ref{fig2}(b). 
We obtain the black solid line 
from the data of $N_{\rm s}$=42 and 36 
as representatives of the data of even $N_{\rm s}$;  
the line indicates $F=-0.03955$. 
The black dotted line is also drawn so that 
the line runs through the origin and 
the data point of $N_{\rm s}$=42. 
Note that the solid and dotted black lines 
approach the data point of the largest $\Delta_{30}/J$ and 
the data point of the smaller $\Delta_{24}/J$. 
In the cases of odd $N_{\rm s}$, 
we obtain the red solid line 
from the data point of $N_{\rm s}$=39 and 
the average of the two data of $N_{\rm s}$=33 
as representatives of the data of odd $N_{\rm s}$;  
the line indicates $F=0.0007$. 
It is worth mentioning that 
the red line goes very near the data point of $N_{\rm s}$=27.  
This fact supports the notion 
that treating the average of the two data 
for $N_{\rm s}=33$ is reasonable. 
Since $F=0.0007$ is quite small, 
if one compares it with the difference between 
the two values of $\Delta_{N_{\rm s}}$ for $N_{\rm s}=33$, 
we can conclude that $F$ is vanishing in the case 
of odd $N_{\rm s}$. 
Although the two values of $F$ 
from even $N_{\rm s}$ and odd $N_{\rm s}$ 
are different from each other, 
the difference comes from the effect that 
higher-order terms appear differently between the cases 
of even $N_{\rm s}$ and odd $N_{\rm s}$.  
The two results from even $N_{\rm s}$ and odd $N_{\rm s}$ 
commonly reveal the gapless feature 
of the two-dimensional system 
shown in Fig.~\ref{fig2}(b).  
We emphasize here that no contradictory points are found 
in our numerical-diagonalization results 
in the viewpoint that the system is gapless. 


In summary, we have studied the spin gap issue of 
the kagome-lattice Heisenberg antiferromagnet 
by the numerical-diagonalization method 
uging the Lanczos algorithm. 
We have successfully obtained 
a finite-size gap of the 42-site cluster; 
we have examined the system-size dependence of gaps 
of all the available clusters up to 42 sites. 
From the results of the analyses from the viewpoints 
of the gapped and gapless cases, 
we here conclude that the system is gapless. 
This conclusion is in agreement with the fact that 
no spin gap is observed experimentally\cite{Mendels}.  
On the other hand, the present conclusion is not 
in agreement with the recent result from the DMRG analysis; 
the discrepancy should be resolved in future studies. 
The consequence of the gaplessness will be followed by 
the examination of whether or not 
the long-range order exists, 
although it is difficult 
to appropriately choose possible candidates 
from among the classical spin arrangements.  
To find a good candidate, 
the spin correlation functions of large systems 
would be a good help.  
To better understand the properties of the system, 
the wave numbers of the wave functions and their degeneracy 
should be examined. 
References \ref{Lecheminant} and \ref{Waldtmann} showed 
part of such results for $N_{\rm s}=27$ and $N_{\rm s}=36$, 
respectively; 
information on larger systems is desired 
in future studies\cite{comment_wave_funct42}. 
Numerical-diagonalization calculations of systems 
as large as possible will provide us 
with a new and better understanding of condensed-matter physics. 

\section*{Acknowledgments}
We wish to thank 
Professors K.~Hida, T.~Tonegawa, S.~Miyashita, 
M.~Imada, and M.~Isoda, 
and Dr.~Y.~Okamoto 
for fruitful discussions. 
This work was partly supported by a Grant-in-Aid (No.~20340096) 
from the Ministry of Education, Culture, Sports, Science 
and Technology of Japan. 
This work was partly supported by 
a Grant-in-Aid (No. 22014012) 
for Scientific Research and Priority Areas 
``Novel States of Matter Induced by Frustration'' 
from the Ministry of Education, Culture, Sports, Science 
and Technology of Japan. 
Nonhybrid thread-parallel calculations 
in the numerical diagonalizations were based 
on TITPACK ver.2, coded by H. Nishimori. 
Part of the computations were performed using 
the facilities of 
the Information Technology Center, Nagoya University;  
Department of Simulation Science, 
National Institute for Fusion Science (NIFS); 
and the Supercomputer Center, 
Institute for Solid State Physics (ISSP), University of Tokyo. 
We also thank Professor
Naoki Kawashima and the staffs of the Supercomputer
Center, ISSP for their support to carry out 
our large-scale parallelized calculations 
in a special job class of the supercomputer of ISSP.



\begin{thebibliography}{99}
\bibitem{comm_performance}
The case of the largest dimension, namely, $N_{\rm s}=42$ and 
$M=0$, has taken approximately 680 seconds per Lanczos step 
in an MPI-parallelized job with 8192 processes 
in SGI Altix ICE 8400EX, ISSP, University of Tokyo. 
The case of $N_{\rm s}=39$ and $M=1/2$ 
has taken about 139 seconds per Lanczos step 
in a hybrid parallel job with 512 processes by 16 threads 
in Hitachi SR16000, Department of Simulation Science, NIFS.
\bibitem{comments2para}
This hybrid code was developed originally 
in the study of the precise estimate of the Haldane gap 
in ref.~\ref{HN_Terai}.
\bibitem{comment_dim}
The record before the present study 
was 159~059~993~400
in the study of the two-dimensional doped Hubbard model 
in refs.~\ref{y_m1} and \ref{y_m2}.
\bibitem{comment_Mendels}
In Fig.~13 of ref.~\ref{Mendels}, 
the linear fitting of a mixture of data of even $N_{\rm s}$ 
and odd $N_{\rm s}$ was carried out in the plot 
of the $1/N_{\rm s}$ dependence of $\Delta_{N_{\rm s}}/J$. 
Although the fitting gives the vanishing gap, 
treating data of even $N_{\rm s}$ and odd $N_{\rm s}$ together 
is not appropriate. 
Actually, one finds in Fig.~\ref{fig3} of the present study 
that the two sequences of data are different from each other.
\bibitem{comm_negative}
A negative $F$ is considered to come 
from the neglected higher-order terms 
with respect to $1/N_{\rm s}^{1/2}$.
\bibitem{comment_wave_funct42}
Unfortunately, there is insufficient memory 
in the supercomputer of ISSP 
to calculate the ground-state wave function 
of the $N_{\rm s}$=42 system at the present time, 
although we have successfully calculated only its energies.
\end{thebibliography}
\end{document}